\documentclass[epj]{svjour}
\usepackage{graphicx,amsmath}
\begin{document}
\title{Thermalization of plastic flow versus stationarity of thermomechanical equilibrium in SGR theory}
\author{Robert S. Hoy}                     
\institute{Department of Physics, University of South Florida, Tampa, FL 33620, USA}

\date{Received: date / Revised version: date}

\abstract{
We discuss issues related to thermalization of plastic flow in the context of soft glassy rheology (SGR) theory.
An apparent problem with the theory in its current form is that the stationarity of thermomechanical equilibrium obtained by requiring that its flow rule satisfy detailed balance in the absence of applied deformation requires plastic flow to be athermal.
This prevents proper application of SGR to small-molecule and polymer glasses where plastic flow is often well-thermalized.
Clearly, one would like to have a SGR-like theory of thermalized plastic flow that satisfies stationarity.
We discuss reasons why such a theory could prove very useful and clarify obstacles that must be overcome in order to develop it.}

%
%
\maketitle
\section{Elucidation of the problem}
\label{sec:intro}

One of the most exciting recent developments in physical mechanics is the discovery that structural glasses are composed of spatially localized, mesoscale \textit{plastic zones} with a wide range of thermodynamic and mechanical stabilities \cite{tsamados09,riggleman10b,manning11,schoenholz14,rodney11,swayamjyoti14,swayamjyoti16,ding14,patinet16}.
Conceptually speaking, materials composed of plastic zones with a wide range of stability mesh very well with modern mean-field plasticity theories such as shear transformation zones (STZ) \cite{falk98,langer04,langer08} and soft glassy rheology (SGR) \cite{sollich97,sollich98,fielding00}.
Such theories are consistent with the recently developed rigorous nonequilibrium thermodynamics of deforming systems \cite{bouchbinder09a,bouchbinder09b,langer12,sollich12,bouchbinder13,fuereder13}, and
have also been supported (at least in part) by recent experiments \cite{pan08,dmowski10,marruzzo13,cubuk17}  showing that glasses are highly elastically heterogeneous.
It is therefore clear that mesoscale plasticity theories taking plastic zones as their fundamental degrees of freedom show great promise as a means to improving our understanding of these systems' mechanical response.

An obstacle currently preventing full realization of this promise is that many theoretical issues related to the \textit{thermalization} of plastic flow (the degree to which the character of systems' plastic deformation is determined by the state of their ``fast'' \cite{bouchbinder09a,bouchbinder09b,langer12}, microscale-vibrational degrees of freedom) remain unresolved \cite{barrat18}.
The reason understanding thermalization is so important is that thermalization controls the configurations plastic zones adopt \textit{after} yielding:\ understanding it allows us to answer the question of how these configurations depend on temperature and the applied strain rate.
This in turn is essential to accurately predicting the rate- and temperature-dependence of systems' mechanical response, which is one of the main aims of plasticity theory \cite{ree55}.
In this paper, we identify thermalization issues pertinent to SGR theory and address their potential resolution.

SGR theory treats systems as being composed of plastic zones with activation energy $\mathcal{U} = k_B T_g u$ and elastic modulus $\mathcal{K}_u$. 
Their configurational energy is  $\mathcal{E}(u,\epsilon^{el}) = -\mathcal{U} + \mathcal{K}_u [\epsilon^{el}]^2/2$ when their elastic strain is $\epsilon^{el}$.\footnote{Here $\mathcal{U} = k_B T_g u$ is the activation energy of \textit{unstrained} zones.  In general their activation energy is $\mathcal{A}(u,\epsilon^{el}) = -\mathcal{E}(u,\epsilon^{el})$.  Note that while SGR theory does not incorporate the quasiuniversal $\mathcal{A} \propto (\delta - |\epsilon^{el}|)^{3/2}$ proportionality identified by more recent work \cite{maloney06,fan14,fan17}, it can in principle be modified to do so.}
For convenience, we will call these entities ``($u,\epsilon^{el}$)-zones.''
Their yielding rate at temperature $T$ is\newline
$\tau^{-1}(u,\epsilon^{el},T) = \tau_0^{-1} \exp[\beta\mathcal{E}(u,\epsilon^{el})]$, where $\tau_0$ is the characteristic microscopic relaxation time of systems' fast degrees of freedom.
The theory adopts a simplified picture of glasses' potential energy landscapes.
Its mean-field treatment ignores the complex saddle-point structure governing transitions between basins on real systems' PELs \cite{maloney06,fan14,fan17}; all zones are assumed to become unstable (have $\tau < \tau_0$) when $\mathcal{E} > 0$.
Plastic zones are assumed to be internally structureless, and their volume $v_0$ is typically taken to be a material-dependent constant (in contrast to STZ theory, in which shear transformation zones are dilute and their density is a dynamical variable within the theory \cite{langer04,langer08}).
Thus it is assumed that the number of admissible ($u,\epsilon^{el}$)-zone configurations depends only on $u$ and is described by a density-of-states function $\rho(u)$, with the functional form of $\rho(u)$ set by systems' microscopic interactions.

Let $w(u,\epsilon^{el})$ be the statistical weight of  ($u,\epsilon^{el}$)-zones.
The SGR flow equation for constant strain rate deformation is \cite{sollich98} 
\begin{equation}
\displaystyle\frac{d w(u,\epsilon^{el})}{d t} = -\dot\epsilon\displaystyle\frac{\partial w(u,\epsilon^{el})}{\partial \epsilon^{el}}   - \displaystyle\frac{ w(u,\epsilon^{el})}{\tau(u,\epsilon^{el},T)}   + f(u,\epsilon^{el},T) \left< \tau^{-1} \right>,
\label{eq:trapflow}
\end{equation}
where $\left< \tau^{-1} \right> = \int \int \frac{w(\tilde{u},\tilde{\epsilon}^{el})}{\tau(\tilde{u},\tilde{\epsilon}^{el},T)} d\tilde\epsilon^{el} d\tilde{u}$ is the average zone yielding rate.
The character of the system's plastic flow is encoded in the functional form of $f(u,\epsilon^{el},T)$.
For the athermal systems for which SGR theory was originally developed (e.g.\ foams, pastes, and slurries \cite{sollich97,sollich98}), $\mathcal{U}$ is typically large compared to $k_B T$ and hence plastic flow is typically nearly athermal.
For this reason as well as for simplicity, SGR's standard formulation \cite{sollich97,sollich98,fielding00} picks $f(u,\epsilon^{el},T) = \rho(u)\Theta(\delta-|\epsilon^{el}|)$, where $\rho(u)$ is the density of single-plastic-zone microstates on the system's energy landscape, $\Theta$ is the Heaviside step function, and $\delta$ is zones' athermal yield strain.
This means that when zones yield, their configurations ($u,\epsilon^{el}$) are reset to new configurations ($u',\epsilon^{el'}$) that are minimally stable [have $\mathcal{E}(u,\epsilon^{el}) < 0$] but otherwise are selected randomly.
In contrast, thermalized SGR \cite{merabia16,hoy17} assumes that the new configurations are fully equilibrated  by systems' fast degrees of freedom.
Specifically, it assumes that fully-thermalized plastic flow is obtained when $f(u,\epsilon^{el},T) = w_{eq}(u,\epsilon^{el},T) \Theta(\delta-|\epsilon^{el}|)$, where $w_{eq}(u,\epsilon^{el},T)$ is  ($u,\epsilon^{el}$)-zones'  equilibrium statistical weight.\footnote{For notational convenience, the $\Theta(\delta-|\epsilon^{el}|)$ terms will be suppressed throughout the rest of this paper except where clarity requires making them explicit.}
Since real glassforming systems thermomechanically equilibrate [i.e.\ their $w(u,\epsilon^{el})$ reach $w_{eq}(u,\epsilon^{el},T)$] over finite timescales only for $T > T_g$, this $f(u,\epsilon^{el},T)$ cannot be a fully accurate representation of glassy ($T < T_g$) plastic flow for finite $\dot\epsilon$.
Nonetheless, it represents an idealized limit worth considering, the complement to the similarly idealized perfectly athermal plastic flow modeled by standard SGR.

Consider the zero-applied-strain-rate case, i.e.\ consider systems that are not being actively deformed.
The SGR flow rule (Eq.\ \ref{eq:trapflow}) becomes
\begin{equation}
\displaystyle\frac{d w(u,\epsilon^{el})}{d t} = -\displaystyle\frac{ w(u,\epsilon^{el})}{\tau(u,\epsilon^{el},T)} + f(u,\epsilon^{el},T) \left< \tau^{-1} \right>.
\label{eq:zerorate}
\end{equation}
Since thermomechanical equilibrium is by definition a stationary state, $w_{eq}(u,\epsilon^{el},T)$ should be a stationary solution to Eq.\ \ref{eq:zerorate}.
For nonthermalized flow, Eq.\ 2 becomes
\begin{equation}
\displaystyle\frac{d w(u,\epsilon^{el})}{d t} = -\displaystyle\frac{ w(u,\epsilon^{el})}{\tau(u,\epsilon^{el},T)} + \rho(u)  \left< \tau^{-1} \right>,
\label{eq:nthss}
\end{equation}
which is just Bouchaud's trap model \cite{bouchaud92,monthus96} with strain degrees of freedom added.
Eq.\ \ref{eq:nthss}'s stationary solution is 
\begin{equation}
w^*(u,\epsilon^{el},T) = \rho(u)  \left< \tau^{-1} \right> \tau(u,\epsilon^{el},T);
\label{eq:wstar}
\end{equation}
just as is the case in the trap model \cite{bouchaud92,monthus96}, systems are stationary when zone statistical weights are proportional to their relaxation times.
The occupation probability of  ($u,\epsilon^{el}$)-zones is $p(u,\epsilon^{el}) = w(u,\epsilon^{el})/\rho(u)$.
Thus the stationary solution $p^*(u,\epsilon^{el},T) = w^*(u,\epsilon^{el},T)/\rho(u)$ satisfies
\footnotesize
\begin{equation}
\displaystyle\frac{p^*(u_2,\epsilon^{el}_2,T)}{p^*(u_1,\epsilon^{el}_1,T)} = \displaystyle\frac{\tau(u_2,\epsilon^{el}_2,T)}{\tau(u_1,\epsilon^{el}_1,T)} = \exp\left( -\beta[ \mathcal{E}(u_2,\epsilon^{el}_2) - \mathcal{E}(u_1,\epsilon^{el}_1) ] \right).
\label{eq:statp1}
\end{equation}
\normalsize
The \textit{equilibrium} occupation probability of these zones is $p_{eq}(u,\epsilon^{el},T) = p_{Boltz}(u,\epsilon^{el},T)/\mathcal{Z}$, where the Boltzmann factor 
\begin{equation}
p_{Boltz}(u,\epsilon^{el},T) = \exp[-\beta(\mathcal{E}(u,\epsilon^{el}) + \alpha^2 k_B T_g)]
\label{eq:pboltz}
\end{equation}
and $\mathcal{Z} = \int \int \rho(u) p_{Boltz}(u,\epsilon^{el},T) d\epsilon^{el} du$ is the partition function.
Here $\mathcal{E}(\alpha^2, 0) = -\alpha^2 k_B T_g$ is the energy of the most stable plastic zones compatible with the given system's microscopic interactions.
Clearly, 
\begin{equation}
\displaystyle\frac{p_{eq}(u_2,\epsilon^{el}_2,T)}{p_{eq}(u_1,\epsilon^{el}_1,T)} = \exp( -\beta[ \mathcal{E}(u_2,\epsilon^{el}_2) - \mathcal{E}(u_1,\epsilon^{el}_1) ]).
\label{eq:pratio}
\end{equation}
Comparing this result to Eq.\ \ref{eq:statp1} shows that thermomechanical equilibrium is stationary [$w^*(u,\epsilon^{el},T) = w_{eq}(u,\epsilon^{el},T)$] 
for $f(u,\epsilon^{el},T) = \rho(u)$ when $\dot\epsilon = 0$.
This is, of course, a desired feature of any plasticity theory and indeed is required for thermodynamic consistency.

Now consider thermalized plastic flow.
The thermalized counterpart to Eq.\ \ref{eq:nthss} is \cite{hoy17}
\begin{equation}
\displaystyle\frac{d w(u,\epsilon^{el})}{d t} = -\displaystyle\frac{ w(u,\epsilon^{el})}{\tau(u,\epsilon^{el},T)} + w_{eq}(u,\epsilon^{el},T)  \left< \tau^{-1} \right>.
\label{eq:thermss}
\end{equation}
The essential feature of thermalized SGR is that new zone configurations are selected with probability equal to the equilibrium statistical weights $w_{eq}(u,\epsilon^{el},T)$.
Indeed, this is what ``thermalized plastic flow'' \textit{means}, in contrast to nonthermalized flow where new zone configurations are selected randomly [i.e.\ with probability $\rho(u)$, independent of $T$].
Unfortunately, $w_{eq}(u,\epsilon^{el},T)$ is \textit{not} a stationary solution of Eq.\ \ref{eq:thermss}.
The initial condition $w(u,\epsilon^{el}) = w_{eq}(u,\epsilon^{el},T)$ gives 
\begin{equation}
\displaystyle\frac{d w(u,\epsilon^{el})}{d t} =  w_{eq}(u,\epsilon^{el},T) \left[  \left< \tau^{-1} \right> - \tau^{-1}(u,\epsilon^{el},T) \right]
\label{eq:eqnonstat}
\end{equation}
at $t = 0$, which produces a net flow from higher-$\mathcal{E}$ into lower-$\mathcal{E}$ zone configurations and leads to spurious ``aging'' away from equilibrium with increasing $t$.

This is obviously a serious flaw, and indicates that the use of $f(u,\epsilon^{el},T) = w_{eq}(u,\epsilon^{el},T)$ in Eq.\ \ref{eq:trapflow} (and in Ref.\ \cite{hoy17}) needs to be challenged. 
Nevertheless, there are many reasons to believe a thermalized SGR theory -- and ideally, a \textit{partially/variably} thermalized SGR theory -- is desirable, and we should not give up the effort to develop one.
Prominent among these reasons is the desire to extend the applicability of SGR theory from the nearly-athermal materials it was originally designed to treat to more-thermal amorphous materials such as metallic, small-molecule, and polymeric glasses  \cite{hoy17,merabia16}.
These materials' postyield response depends very strongly on $T$, e.g.\ flow stresses at fixed $\dot\epsilon$ tend to scale approximately as ($1-T/T_g$) \cite{schuh07,roth16}, yet athermal plastic flow by its very nature tends to drive systems to a point on their energy landscapes that depends only weakly on $T$, especially at larger $\dot\epsilon$.
To further illustrate why developing a properly thermalized version of SGR theory is a worthy goal, we will consider a specific example that corresponds to such systems.

\section{Effects of the degree of plastic flow thermalization}

\begin{figure}
\resizebox{.445\textwidth}{!}{%
  \includegraphics{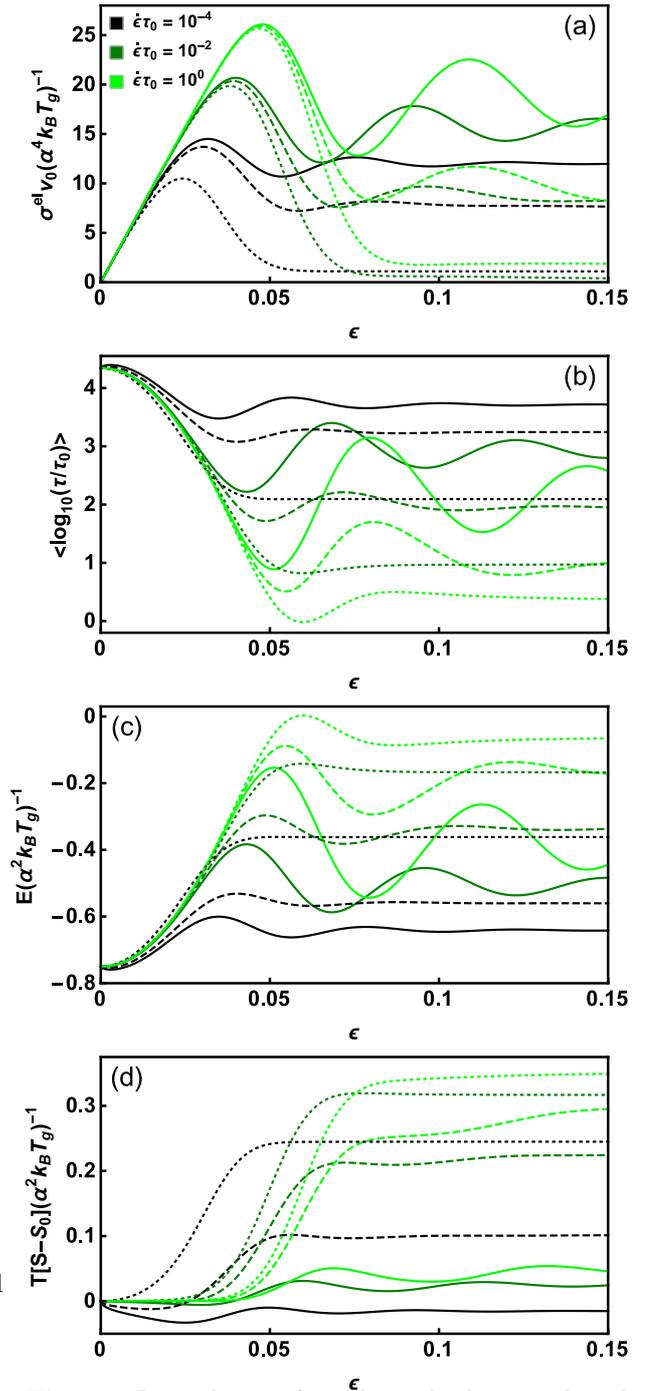}
}
\caption{Dependence of mechanical, dynamical and thermodynamical response of $\alpha^2 = 10$ systems (at $T/T_g = 3/4$) on the character of their plastic flow.
Solid and dotted curves respectively indicate results for fully thermalized flow  [$f(u,\epsilon^{el},T) = w_{eq}(u,\epsilon^{el},T)$] and nonthermalized flow [$f(u,\epsilon^{el},T) = \rho(u)$].
Dashed curves indicate results for half-thermalized flow [$f(u) \propto \sqrt{\rho(u)w_{eq}(u,\epsilon^{el},T)}$; Eq.\ \ref{eq:halftherm}].
Black, dark green, and bright green curves respectively indicate $\dot\epsilon\tau_0 = 10^{-4},\ 10^{-2},\ \rm{and}\ 1$.  
Panel (a):\ elastic stress $\sigma^{el}(\epsilon)$.
Panel (b):\ logarithmically averaged zone relaxation time $\left< \log_{10}(\tau(\epsilon)/\tau_0) \right>$. 
Panels (c-d):\ configurational energy density $E(\epsilon)$ and temperature$\times$entropy density $TS(\epsilon)$.  All energy densities are scaled by $\alpha^2 k_B T_g/v_0$, where $v_0$ is the average volume of a plastic zone; here we take $v_0 = 1$.}
\label{fig:1}       
\end{figure}

\begin{figure}
\resizebox{.445\textwidth}{!}{
  \includegraphics{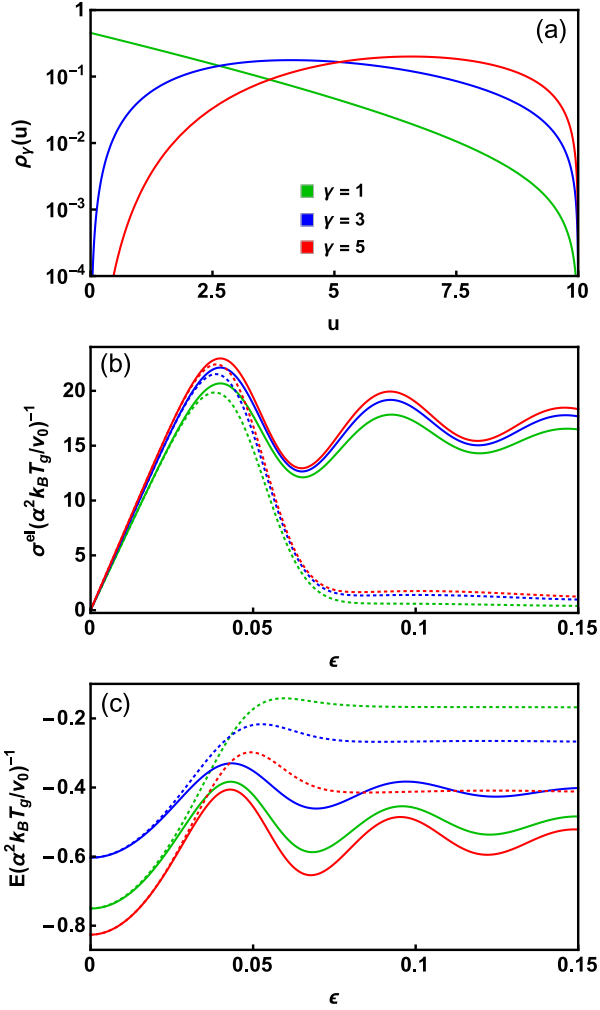}
}
\caption{Dependence of the response  to deformation at moderate strain rate ($\dot\epsilon\tau_0 = 10^{-2}$) and temperature ($T/T_g = 3/4$) of $\alpha^2 = 10$ systems on the shape of their energy landscapes [$\rho_{\gamma}(u)$:\ Eq.\ \ref{eq:rho1}].
Green, blue and red curves respectively show results for $\gamma = 1,\ 3,\ \rm{and}\ 5$, while solid and dashed curves respectively indicate results for thermalized and nonthermalized plastic flow.}
\label{fig:2}       
\end{figure}

Figure \ref{fig:1} shows results for $\alpha^2 = 10$ systems deformed at three characteristic strain rates: low ($\dot\epsilon\tau_0 = 10^{-4}$), moderate ($\dot\epsilon\tau_0 = 10^{-2}$), and  high ($\dot\epsilon\tau_0 = 1$).
This choice of $\alpha$ is motivated by recent soft-spot studies \cite{rodney11,swayamjyoti14,swayamjyoti16} suggesting $\alpha^2 \simeq 10$ for metallic glasses; values for polymeric glasses should be similar \cite{roth16}.
The value of each macroscopic physical quantity $\left<\zeta(\epsilon)\right>$ is given by \cite{fuereder13,hoy17} \footnote{The numerical methods used to integrate Eq.\ \ref{eq:trapflow} and calculate the thermodynamic quantities shown in Figs.\ \ref{fig:1}-\ref{fig:2} are the same as those discussed in Ref.\ \cite{hoy17}.}
\begin{equation}
\left<\zeta  (\epsilon) \right> = \displaystyle\int_0^{\alpha^2} \displaystyle\int_{-\delta}^{\delta+\epsilon} \zeta(u,\epsilon^{el}) w(u,\epsilon^{el}) d\epsilon^{el} du.
\label{eq:thermoavg}
\end{equation}
The bounds of the inner integral in Eq.\ \ref{eq:thermoavg} reflect the SGR-theoretic convention \cite{sollich98} that new zone configurations must be stable, i.e.\ have $\mathcal{E}(u,\epsilon^{el}) \leq 0$.
Zones' spring constants $\mathcal{K}_u =  2\mathcal{U}/\delta^2 \equiv 2k_B T_g u/\delta^2$ are chosen so that zones are stable for $|\epsilon^{el}| \leq \delta$; here we present results for $\delta = .05$.
As in Ref.\ \cite{hoy17}, we employ an idealized, highly aged initial condition wherein systems have reached thermomechanical equilibrium [$w(u,\epsilon^{el})_{t = 0} = w_{eq}(u,\epsilon^{el},T)$] and then numerically integrate Eq.\ \ref{eq:trapflow} forward in time for various $f(u,\epsilon^{el},T)$.

Panel (a) shows systems' stress-strain curves.
As is typical in SGR-theoretic studies \cite{sollich97,sollich98,fielding00}, we focus on the elastic component $\sigma^{el}(\epsilon)$.
Thermalized flow produces strain softening behavior that is consistent with that observed in metallic and polymeric glasses \cite{schuh07,roth16}.
The ratio of $\sigma^{flow}/\sigma^y$ of the steady-state plastic flow stress $\sigma^{flow}$ to the yield stress $\sigma^y$ increases with decreasing strain rate and approaches unity in the $\dot\epsilon\tau_0 \ll 1$ limit.
In contrast, nonthermalized flow produces an unrealistically large degree of strain softening, i.e.\ it produces unrealistically small $\sigma^{flow}/\sigma^y$.
Thermalized flow also produces  stress oscillations (the postyield ``stress undershoots'') that increase in magnitude with increasing $\dot\epsilon \tau_0$ in a manner similar to that observed in experiments on ductile bulk metallic glasses \cite{lu03,johnson02}.
Critically, these undershoots are much smaller \cite{fielding00} for nonthermalized flow, and have not (to the best of our knowledge) previously been predicted or explained by any other mesoscale or microscale plasticity theory.
The SGR-theoretic explanation of the undershoots is as follows:
Many zones have characteristic yield strains that are close to the macroscopic yield strain $\epsilon^y$.
Upon yielding at $\epsilon \simeq \epsilon^y$, thermalized plastic flow tends to reset these zones back to lower-$\mathcal{E}$ configurations, i.e.\ back to small values of $\epsilon^{el}$ and $\sigma_{el}$.
Stress then increases again as deformation continues.
This process can repeat a few times (albeit with diminishing magnitude) before steady-state flow is achieved.
Thermally activated yielding smears out this process, causing the maximum magnitude of undershoots to decrease with decreasing $\dot\epsilon\tau_0$.
The unrealistically large stress oscillations at the largest strain rate considered here probably just indicate that plastic flow at $\dot\epsilon\tau_0 = 1$ \textit{cannot} be fully thermalized \cite{bouchbinder09a,bouchbinder09b,langer12}.

Panel (b) shows results for the average zone relaxation time.
For all systems, $\left< \tau (\epsilon) \right>$ passes through a minimum at $\epsilon \simeq \epsilon^y$ before finally reaching a steady-state value $\left< \tau \right>^{flow}$ indicating perfect plasticity for $\epsilon >\sim \epsilon^{flow}$.
The much larger values of $\left< \tau \right>^{flow}$ for thermalized plasticity are consistent with its much larger $\sigma^{flow}$.
Thermalized flow also produces oscillations in $\left< \tau (\epsilon) \right>$ for $\epsilon^y <\sim \epsilon < \epsilon^{flow}$ that directly correspond to the abovementioned oscillations in $\sigma^{el}(\epsilon)$.
All of these effects are associated with the fact that thermalized flow leaves newly reset zones lower on their energy landscapes. 
Since predicting how deformed systems' (inherently heterogeneous \cite{lee09,bending14}) relaxation dynamics evolve with strain is extremely important to understanding the mechanics of thermal glasses, it is clearly desirable to develop a solid understanding of such effects within SGR.
Unfortunately, this is where the spurious nonstationarity of the present thermalized flow law (Eq.\ \ref{eq:thermss}) shows up.
For the lowest strain rate [$\dot\epsilon\tau_0 = 10^{-4}$], $\left< \tau(\epsilon) \right>$ actually initially increases with increasing strain because spurious aging initially overwhelms the tendency of applied deformation to increase $\left< \mathcal{E} \right>$ and decrease $\left< \tau \right>$. 
Ref.\ \cite{hoy17} failed to identify this issue because it did not examine low strain rates in detail.

Panels (c-d) respectively show results for systems' configurational energy and entropy densities $E = \left<\mathcal{E}\right>$ and $S = -k_B \left< \ln(p) \right>$.
These  clearly show how much further nonthermalized flow drives systems up their energy landscapes.
Since $\rho(u) \propto \exp(-u/\alpha)$ \cite{sollich98,hoy17}, nonthermalized plastic flow preferentially creates zones with smaller $\mathcal{U}$.
Such zones cannot sustain large stresses before yielding again.
This is why the $\sigma^{flow}/\sigma^y$ values for nonthermalized flow are unrealistically low.
Higher strain rates dramatically increase steady-state values of both $E$ and $S$ for both thermalized and nonthermalized flow, but to different degrees.
Spurious aging effects show up again here, in the form of the decreases in energy and entropy at small strains for $\dot\epsilon\tau_0 = 10^{-4}$. 
However, it is important to note that it is only at low strain rates that the nonstationarity of the initial condition ($w = w_{eq}$) under thermalized flow significantly affects any of  the results presented in Figure \ref{fig:1}.
For $\dot\epsilon\tau_0 >\sim 10^{-2}$, the differences between thermalized and nonthermalized response arise primarily because nonthermalized resets preferentially populate the upper regions of systems' energy landscapes.
This will remain true for modified $f(u,\epsilon^{el},T)$ that restore the stationarity of $w = w_{eq}$ for $\dot\epsilon = 0$.

Fig.\ \ref{fig:1} also presents results for an ad hoc model of plastic flow that is intermediate between nonthermalized and fully thermalized.
Dashed curves show results for ``half-thermalized'' flow with
\begin{equation}
f(u,\epsilon^{el}, T) = \displaystyle\frac{\sqrt{\rho(u)w_{eq}(u,\epsilon^{el},T)}\Theta(\delta-|\epsilon^{el}|)}{\displaystyle\int_0^{\alpha^2} \displaystyle\int_{-\delta}^{\delta} \sqrt{\rho(u)w_{eq}(u,\epsilon^{el},T)} d\epsilon^{el} du}.
\label{eq:halftherm}
\end{equation}
In all cases, results for half-thermalized plastic flow lie intermediate between those for nonthermalized and fully-thermalized flow.
Note that Eq.\ \ref{eq:halftherm} represents one special case of a more general flow rule
\begin{equation}
f_\theta(u,\epsilon^{el}, T) = \displaystyle\frac{\rho(u)^{1-\theta}w_{eq}^\theta(u,\epsilon^{el},T)\Theta(\delta-|\epsilon^{el}|)}{\displaystyle\int_0^{\alpha^2} \displaystyle\int_{-\delta}^{\delta} \rho(u)^{1-\theta} w_{eq}^\theta(u,\epsilon^{el},T) d\epsilon^{el} du},
\label{eq:partialtherm}
\end{equation}
where the thermalization parameter $\theta$ satisfies $0 \leq \theta \leq 1$.
Fig.\ \ref{fig:1} shows results for $\theta = 0,\ 1/2,\ \rm{and}\ 1$.
Results for other $\theta$ show a continuous crossover from the nonthermalized to the thermalized limits as $\theta$ increases.
$\theta$ should in principle be predictable from knowledge of systems' microscopic interactions and depend on experimental conditions.
For example, since the characteristic timescale for thermalization is $\tau_0$, one expects \cite{bouchbinder09a,bouchbinder09b,langer12} that experiments conducted at very high strain rates ($\dot\epsilon\tau_0 >\sim 1$) have $\theta \ll 1$ whereas experiments conducted at very low strain rates ($\dot\epsilon\tau_0 \ll 1$) would have $\theta \simeq 1$. 
Developing a rigorous theory of the $\theta$ parameter would presumably require developing a rigorous theory of the glass transition, and hence remains an unrealized goal.
Nonetheless, the crossover between nonthermalized and fully-thermalized plastic flow is worth exploring in more detail because plastic flow in experiments on real glasses presumably always lies in between these idealized limits.

One could argue that the abovementioned differences between the responses for thermalized and nonthermalized plastic flow are misleading in the context of thermal glasses because these systems' constituents have attractive interactions and thus should possess few low-$\mathcal{U}$ plastic zone configurations  \cite{rodney11,swayamjyoti14,swayamjyoti16}, whereas the exponential $\rho(u) \propto \exp(-u/\alpha)$ energy landscape typically employed in both the trap model and SGR has many such configurations.
If this argument is valid, nonthermalized flow's tendency to populate the upper regions of systems' energy landscapes is less relevant.
A generic form for $\rho(u)$ that imposes a variable low-$\mathcal{U}$ cutoff is
\begin{equation}
\rho_\gamma(u) =  \displaystyle\frac{C_{\gamma}(\alpha)}{2\alpha\delta}  \left(\displaystyle\frac{u}{\alpha}\right)^{\gamma-1} \left( 1 - \displaystyle\frac{u}{\alpha^2}\right)  \exp\left(-\displaystyle\frac{u}{\alpha} \right),
\label{eq:rho1}
\end{equation}
where $C_{\gamma}(\alpha)$ is the normalization constant satisfying\newline
$\int_0^{\alpha^2} \rho_\gamma(u) du = 1$.
$\gamma = 1$ systems have the standard cutoff-trap-model  energy landscape wherein $\rho(u)$ is maximized at $u = 0$; they have an abundance of very shallow minima corresponding to very soft, easily deformed plastic zones.
For $\gamma > 1$, the $(u/\alpha)^{\gamma-1}$ proportionality in Eq.\ \ref{eq:rho1} imposes a low-$u$ cutoff; systems have a vanishing number of very shallow energy minima [$\lim_{u\to 0} \rho_\gamma(u) \propto u^{\gamma-1} \rightarrow 0$].
Such low-$u$ cutoffs are physically reasonable for thermal glasses because very shallow minima will also be very soft and will often be mechanically unstable at finite $T$ \cite{riggleman10b,manning11,fan14,fan17}.
Increasing $\gamma$ also increases the relative influence of very deep energy minima (mechanically ``hard'' zones \cite{tsamados09,riggleman10b,manning11,schoenholz14,rodney11,swayamjyoti14,swayamjyoti16,ding14,patinet16}).
Fig.\ \ref{fig:1} showed the response for this form of $\rho(u)$ with $\gamma = 1$.

Figure \ref{fig:2} shows the $\gamma$-dependent differences in the response that are most relevant to our present purpose.
Panel (a) illustrates the very different shapes of the various $\rho_\gamma(u)$.
Panels (b-c) contrast $\dot\epsilon\tau_0 = 10^{-2}$ results for $\gamma = 1,\ 3,\ \rm{and}\ 5$.
They show that the differences in the responses for thermalized vs.\ nonthermalized flow depend quantatively but not qualitatively on $\gamma$.
In particular, nonthermalized flow produces excessive strain softening and drives systems unrealistically far up their energy landscapes for all three values of $\gamma$.
We conclude that the above-identified issues with applying nonthermalized SGR to model thermal glasses are not an artifact of a poorly chosen $\rho(u)$.

\section{Why the most obvious strategies for restoring stationarity fail}

SGR is a very promising framework for studying effects like those discussed above because its treatment of systems as ensembles of plastic zones with broadly distributed activation energies and elastic constants meshes straightforwardly with the findings of the recent soft-spot studies \cite{tsamados09,riggleman10b,manning11,schoenholz14,rodney11,swayamjyoti14,swayamjyoti16,ding14,patinet16}.
Since thermodynamic consistency is an essential feature for a generally valid plasticity theory \cite{bouchbinder09a,bouchbinder09b}, it is clearly highly desirable to be able to treat thermalized plastic flow within a version of SGR that satisfies our desired stationarity condition.
The question is therefore:\ how can we modify the SGR flow law to produce thermalized plastic flow that has the correct stationary solution $w^*(u,\epsilon^{el},T)  =  w_{eq}(u,\epsilon^{el},T)$?

The most obvious strategy is to invoke the condition that the correct $f(u,\epsilon^{el},T)$ should satisfy detailed balance.
Consider two plastic zone configurations: $A$ and $B$.
Detailed balance requires
\begin{equation}
w_{eq}^A \mathcal R_{A \to B} = w_{eq}^B \mathcal R_{B \to A}, 
\label{eq:fromdb1}
\end{equation}
where $\mathcal{R}_{A \to B}$ and $\mathcal{R}_{B \to A}$ are respectively the rates of $A\to B$ and $B\to A$ transitions and are respectively equal to $\tau^{-1}(u_A, \epsilon^{el}_A, T) f(u_B, \epsilon^{el}_B, T)$ and  $\tau^{-1}(u_B, \epsilon^{el}_B, T) f(u_A, \epsilon^{el}_A, T)$.
Plugging these identities in, Eq.\ \ref{eq:fromdb1} becomes
\begin{equation}
w_{eq}^A \tau^{-1}_A f(u_B, \epsilon^{el}_B, T) =  w_{eq}^B \tau^{-1}_B f(u_A, \epsilon^{el}_A, T).
\end{equation}
Rearranging terms, this condition becomes
\begin{equation}
\displaystyle\frac{ f(u_B, \epsilon^{el}_B, T) } { f(u_A, \epsilon^{el}_A, T) } = \displaystyle\frac{  w_{eq}^B \tau^{-1}_B }{ w_{eq}^A \tau^{-1}_A } = \displaystyle\frac{ \rho(u_B) }{ \rho(u_A) }.
\end{equation}
Annoyingly, this implies that $f(u,\epsilon^{el},T) = \rho(u)$ and therefore that plastic flow described by rules with the form of Eq.\ \ref{eq:trapflow} and satisfying detailed balance must \textit{necessarily} be athermal!

An alternative strategy is to assume ($u,\epsilon^{el}$)-zones are created with the characteristic rate $\tau^{-1}(u,\epsilon^{el},T)$, and set 
\begin{equation}
f(u,\epsilon^{el},T) =  \displaystyle\frac{w_{eq}(u,\epsilon^{el},T)}{\left< \tau^{-1} \right>\tau(u,\epsilon^{el},T)},
\label{eq:badf1}
\end{equation}
so that Eq.\ \ref{eq:thermss} becomes
\begin{equation}
\displaystyle\frac{d w(u,\epsilon^{el})}{d t} = -\displaystyle\frac{w(u,\epsilon^{el})}{\tau(u,\epsilon^{el},T)} + \displaystyle\frac{w_{eq}(u,\epsilon^{el},T)}{\tau(u,\epsilon^{el},T)}.
\label{eq:badmod1}
\end{equation}
This modification of $f$ favors creation of higher-$\mathcal{E}$ zones and eliminates the spurious aging.
It obviously satisfies our desired stationarity condition:\ for all $u$ and $\epsilon^{el}$, $w$ decays towards $w_{eq}$ at rate $\tau^{-1}(u,\epsilon^{el},T)$.
Unfortunately, it achieves stationarity at the expense of thermalization.
The new-zone-creation term on the right hand side of Eq.\ \ref{eq:badmod1} is equal to $\rho(u) \left< \tau^{-1} \right>$, and thus Eq.\ \ref{eq:badmod1} is equivalent to Eq.\ \ref{eq:nthss}.
In other words, assuming zones with higher configurational energies are created faster [owing to the $\tau^{-1}(u,\epsilon^{el},T)$ proportionality of the zone creation term in Eq.\ \ref{eq:badf1}] dethermalizes flow and defeats our purpose.

Another obvious alternative is to require only the weaker (non-detailed) balance condition that the creation rate of ($u,\epsilon^{el}$)-zones is equal to their yielding rate.
Unfortunately, this does not help.
Within SGR theory, this strategy  just corresponds to replacing configuration $B$ with the thermal reservoir $R$.
Then Eq.\ \ref{eq:fromdb1} becomes $w_{eq}^A \tau^{-1}_A = f(u_A, \epsilon^{el}_A, T) \left< \tau^{-1} \right>$, which again leads to\newline
$f(u, \epsilon^{el}, T) = w_{eq}/[ \left< \tau^{-1} \right> \tau(u,\epsilon^{el},T)]$, i.e.\ to Eq.\ \ref{eq:badf1}.
This in turn highlights an additional conceptual problem.
The equivalence of Eqs.\ \ref{eq:nthss} and \ref{eq:badmod1} indicates that (within SGR theory) athermal plastic flow is obtained when  \textit{and only when} both yielding and creation of ($u,\epsilon^{el}$)-zones occur with the same characteristic rate $\tau^{-1}(u,\epsilon^{el},T)$.
Both physical intuition and theoretical precedent suggest that these rates should not, in general, be identical.
For example, in the well-studied Eyring model \cite{ree55}, the corresponding processes are forward and reverse activation over a barrier in a stress-tilted energy landscape, and reverse activation is exponentially slower.

Yet another potential strategy is to replace Eq.\ \ref{eq:trapflow} with a Chapman-Kolmogorov master-equation style formulation 
\begin{equation}
\begin{array}{l}
\displaystyle\frac{d w(u,\epsilon^{el})}{d t} = -\dot\epsilon\displaystyle\frac{\partial w(u,\epsilon^{el})}{\partial \epsilon^{el}}   - \displaystyle\frac{ w(u,\epsilon^{el})}{\tau(u,\epsilon^{el},T)} \\
\\
\ \ \ +  \displaystyle\int \displaystyle\int g(u,\epsilon^{el}, \tilde{u},\tilde{\epsilon}^{el}; \dot\epsilon, T) \displaystyle\frac{ w(\tilde{u},\tilde{\epsilon}^{el})}{\tau(\tilde{u},\tilde{\epsilon}^{el},T)}d\tilde{\epsilon}^{el} d\tilde{u},
\end{array}
\label{eq:CKflow}
\end{equation}
where $g(u,\epsilon^{el}, \tilde{u},\tilde{\epsilon}^{el}; \dot\epsilon, T)$ is the transition kernel governing ($\tilde{u},\tilde{\epsilon}^{el}$)$\rightarrow$($u,\epsilon^{el}$) zone resets when the applied strain rate is $\dot\epsilon$ and the temperature is $T$.
Recent simulations \cite{fan14,fan17} employing activation-relaxation techniques \cite{barkema96} have suggested that the activated relaxations in glasses are ``memoryless'', i.e.\ the metabasins occupied after local relaxation events are uncorrelated with those occupied prior to activation, and also that activation over saddle points on systems' potential energy landscapes essentially ``melts'' them.
Naively, this would imply that $g(u,\epsilon^{el},\tilde{u},\tilde{\epsilon}^{el}; 0, T)$ depends only on $u$, $\epsilon^{el}$, and $T$, with the $T$-dependence arising from the fact that finite-$T$ plastic arrangements are not limited to direct traversals over saddle points.
However, Refs.\ \cite{fan14,fan17} did not consider rate-dependence, and the functional form $g(u,\epsilon^{el}, \tilde{u},\tilde{\epsilon}^{el}; \dot\epsilon, T)$ should take is far from obvious.

\section{Discussion and Conclusions}

In conclusion, we have shown that both the standard nonthermalized \cite{sollich97,sollich98} and recently proposed thermalized \cite{hoy17} versions of SGR theory have limitations related to the assumptions implicit in their selection rules for new plastic zone configurations that prevent them from being optimally applied to ``thermal'' (metallic, small-molecule, and polymer) glasses.
For systems with low and moderate $\alpha$, the nonthermalized version's implicit assumption that $(u,\epsilon^{el}$)-zones are created at the same rate at which they yield [i.e.\ $\tau^{-1}(u,\epsilon^{el},T)$] drives systems too far up their energy landscapes.
Comparing the dotted and solid curves in Fig.\ \ref{fig:1} clearly illustrates that the latter are more representative of the typical behavior of these glasses \cite{schuh07,roth16}, and Fig.\ \ref{fig:2} shows that this result does not depend strongly on the shape of the energy landscape.
On the other hand, the thermalized version's assumption that all new zones are created at the same (effectively infinite) rate makes thermomechanical equilibrium nonstationary because it precludes detailed balance.

One could argue that SGR theory has been largely superseded by more recently developed plasticity theories which treat cooperative effects such as interzone elastic coupling, mechanical facilitation, and stress diffusion \cite{hebraud98,johnson05,dequidt16,bouchaud16}, and that the issues we have identified above further indicate that it should be abandoned. 
On the other hand, SGR has three redeeming features that indicate it should instead be further developed: (i) it transparently depicts systems as ensembles of plastic zones; (ii) mean-field versions of facilitation and/or stress diffusion can be added to it with little difficulty; and (iii) its flow law (Eq.\ \ref{eq:trapflow}) is amenable to continuously variable thermalization. 
Feature (iii) is especially appealing in light of the many recent studies \cite{langer12,merabia16,hoy17,derlet14,agoritsas15} showing how sensitively the character of plastic flow depends on the degree to which it is thermalized.
Since variable thermalization is intimately connected to variable ``frustration'' (the tendency of newly reset plastic zone configurations to be ``close to'' the just-yielded configurations on systems' potential energy landscapes \cite{sollich98,fan14,fan17}), and these connections also can be systematically explored within SGR-like theories using flow laws like Eq.\ \ref{eq:CKflow}, it seems worthwhile to at least attempt to explore the abovementioned effects within a SGR-like theory obeying a proper stationarity condition.
This presents a challenge for the plasticity-theory community:\ can we resolve the issues identified herein through either a suitable modification of $f(u,\epsilon^{el},T)$ or a more general modification of Eq.\ \ref{eq:trapflow} that preserves the essential spirit of SGR theory?

This material is based upon work supported by the National Science Foundation under Grant No.\ DMR-1555242.
Samy Merabia and David M.\ Rogers provided helpful discussions.

%

\end{document}